\documentclass[letterpaper]{article} % DO NOT CHANGE THIS
\usepackage{aaai25}
\usepackage{times}  % DO NOT CHANGE THIS
\usepackage{helvet}  % DO NOT CHANGE THIS
\usepackage{courier}  % DO NOT CHANGE THIS
\usepackage[hyphens]{url}  % DO NOT CHANGE THIS
\usepackage{graphicx} % DO NOT CHANGE THIS
\def\UrlFont{\rm}  % DO NOT CHANGE THIS
\usepackage{natbib}  % DO NOT CHANGE THIS AND DO NOT ADD ANY OPTIONS TO IT
\usepackage{caption} % DO NOT CHANGE THIS AND DO NOT ADD ANY OPTIONS TO IT
\usepackage{longtable}
\usepackage{booktabs}
\usepackage{cite}
\usepackage{amsmath,amssymb,amsfonts}
\usepackage{algorithmic}
\usepackage{graphicx}
\usepackage{textcomp}
\usepackage{xcolor}
\usepackage{tabularray}
\usepackage{graphicx}
\usepackage{caption}
\usepackage{tabularx}
\usepackage{multirow}

\usepackage{algorithm}
\usepackage{algorithmic}
\usepackage{tikz}
\usepackage{eso-pic}

\usepackage{newfloat}
\usepackage{listings}
\DeclareCaptionStyle{ruled}{labelfont=normalfont,labelsep=colon,strut=off} % DO NOT CHANGE THIS
\floatstyle{ruled}
\newfloat{listing}{tb}{lst}{}
\floatname{listing}{Listing}
\title{PRAC3 (Privacy, Reputation, Accountability, Consent, Credit, Compensation): \\Long Tailed Risks of Voice Actors in AI Data-Economy
}

\author {
 Tanusree Sharma\textsuperscript{\rm 1},
    Yihao Zhou\textsuperscript{\rm 1},
    Visar Berisha\textsuperscript{\rm 2}
}
\affiliations {
    \textsuperscript{\rm 1}Pennsylvania State University \thanks{Corresponding author: Tanusree Sharma}\\
    \textsuperscript{\rm 2}Arizona State University\\
    \textsuperscript{\rm 1} tanusree.sharma@psu.edu
}

\usepackage{bibentry}
\begin{document}

\maketitle

\begin{abstract}
Early large-scale audio datasets, such as LibriSpeech, were built with hundreds of individual contributors whose voices were instrumental in the development of speech technologies, including audiobooks and voice assistants. Yet, a decade later, these same contributions have exposed voice actors to a range of risks. While existing ethical frameworks emphasize Consent, Credit, and Compensation (C³), they do not adequately address the emergent risks involving vocal identities that are increasingly decoupled from context, authorship, and control. Drawing on qualitative interviews with 20 professional voice actors, this paper reveals how synthetic replication of voice without clear provenance or enforceable constraints exposes individuals to both reputational and security threats.

Beyond reputational harm, such as re-purposing voice data in erotic content, offensive political messaging, and meme culture, we document concerns about accountability breakdowns when their voice is leveraged to clone voices that are deployed in high-stakes scenarios such as financial fraud, misinformation campaigns, or impersonation scams. In such cases, actors face social and legal fallout without recourse, while very few of them have a legal representative or union protection. To make sense of these shifting dynamics, we introduce the PRAC³ framework - an expansion of C³ that foregrounds Privacy, Reputation, Accountability, Consent, Credit, and Compensation as interdependent pillars of data used in the synthetic voice economy. This framework captures how privacy risks are amplified through non-consensual training, how reputational harm arises from decontextualized deployment, and how accountability can be reimagined AI Data ecosystems. We argue that voice, as both a biometric identifier and creative labor, demands governance models that restore creator agency, ensure traceability, and establish enforceable boundaries for ethical reuse.
\end{abstract}

\section{Introduction}
\label{sec:intro}
 Data sharing has long been a contested domain between individual contributors, professionals, and data controllers. Individuals or groups contribute data either deliberatively, whether in pursuit of social value, to receive financial compensation, or as part of their primary profession~\cite{allen1999privacy, lane2014privacy, board2015sharing, godard2003data}. These contributions are influenced by a variety of motivations, such as financial incentives from industry~\cite{massiceti2021orbit} and academia~\cite{tseng2024biv, sharma2023disability}, creative expression and social interaction on online platforms~\cite{andalibi2017sensitive, lee2016predicting}, participation in public-interest initiatives like Mozilla Common Voice~\cite{ardila2019common}, LibriSpeech~\cite{panayotov2015librispeech} and AESDD~\cite{data1}. The European Commission estimates that data sharing has the potential to save billions of euros~\cite{EU1}. Shared data are typically governed by various licensing frameworks, including Creative Commons~\cite{lin2014microsoft, russakovsky2015imagenet}, Open Data Commons (ODC)~\cite{miller2008open}, GNU General Public License (GPL)~\cite{license1989gnu}.

Despite these practices, recent legal developments indicate increasing friction between AI companies and creative workers~\cite{nbcnewsScarlettJohansson, gero2025creative, shan2023glaze, kyi2025governance}. In 2024, a YouTube creator initiated a lawsuit against OpenAI, arguing the company transcribed millions of hours of video content to train models like ChatGPT without consent~\cite{hollywoodreporterYouTubeCreators}. Likewise, voice actor Bev Standing filed legal action against TikTok regarding the unauthorized use of her voice in its text-to-speech feature~\cite{bbcActorSues}. Although companies like OpenAI reported that their training datasets consist of publicly available resources~\cite{openaiChatGPTFoundation}, public access does not automatically grant legal or ethical approval for such usage~\cite{gao2024documenting, hoffman2015citizen}. Therefore, creative workers contend that their materials were gathered without authorization, credit, or financial remuneration, resulting in several lawsuits, purported violations of terms of service, and coordinated protests~\cite{youtube, npr}. These cases reflect broader concerns among creative professionals about the unconsented use of their work for AI training, particularly when such work is copyrighted or personally attributable.

While prior work reasonably explored the risk of creative workers, particularly visual artists and writers, research on voice professionals remains limited. Much of the current discourse focuses on fairness in compensation, credit, and consent~\cite{nbcnewsScarlettJohansson, gero2025creative, shan2023glaze, kyi2025governance, gero2025creative} and frames Gen AI as a collaboration tool for creativity~\cite{liu2024ai, chakrabarty2024creativity}. In contrast, voice actors and narrators carry a unique intersection of creative labor and biometric vulnerability. Unlike textual or visual data, voice is not only expressive but also biometric, and it is uniquely identifiable to a person~\cite{aleksic2006audio}. Thus, voice contributors are prone to a wide range of harms, including unauthorized cloning, impersonation, reputational damage, and identity theft~\cite{hutiri2024not,iappIAPP}; however these risks have received little systematic attention.

Moreover, voice actors and contributors played a foundational role in the development of speech technologies~\cite{prahallad2010semi, prahallad2010learning, tauberer2010learning}. A notable innovation was the early large-scale audio datasets, LibriSpeech, derived from thousands of contributions to LibriVox and other public domain audiobook platforms, underpinned early breakthroughs in automatic speech recognition and the voice assistants we use today~\cite{van2007online, kearns2014librivox, prahallad2010automatic, panayotov2015librispeech}. These contributions, originally made in the spirit of open knowledge and accessibility, have since been repurposed into commercial AI pipelines often without consent, attribution, or safeguards~\cite{bbcActorSues}. A decade later, these same contributions have exposed voice actors to a range of harms and may automate, devalue, or displace the very actors who created it.
Yet, despite their centrality to the voice technology landscape, voice actors remain underrepresented in discussions of data labor and AI ethics and risk pertaining to the voiceprint (both a personal and professional tool for voice actors). To address this urgent gap, this study investigates how professional voice actors perceive, negotiate, and respond to risk in the generative AI landscape. 

%Despite their central role in the development of AI technologies, voice actors remain underrepresented in discussions about data labor, algorithmic accountability, and biometric risk. In an era of proliferating generative voice technologies, voice actors face mounting risks related not limited to Consent, Credit, Compensation, but expanded to unauthorized appropriation, replication, and monetization of their vocal identities, cloning with an intent for fun, memes leading to reputational risks, usage for cloning leading to fraud and scam and accountability risk—the core asset of their creative and professional practice.  In the creative domain, we know a little about voice and voice workers.  To address this gap, this study investigates how voice actors navigate risks in a rapidly shifting AI landscape

\begin{enumerate}
 \item \textbf{RQ1:} In what ways do voice actors recognize, interpret, and negotiate risk when engaging with digital platforms, clients, and publishers, given the rise of generative AI?
    \item \textbf{RQ2:} How do voice actors perceive the long-term risks associated with voice data? 
    \item \textbf{RQ3:} How do voice actors’ perceptions and lived experiences of risks contribute to forming threat models in assessing risk over time? 
 \end{enumerate}

%\tanusree{add three bullet point of unique findings to this work, not common with the other type of creative work}
To answer these questions, we interviewed a total of 20 voice actors at different stages of their careers and in different work modes (e.g., freelancers, contract workers, and studio owners). We found that voice actors face unique challenges that are different from other creative workers, including:  \textbf{1) Biometric Identity Risks}: Voice data combines creative work with biometric identity, thus exposing voice actors to unique risks of unauthorized cloning, identity theft, and reputational harm when their recordings are misused in unauthorized illegal contexts.
\textbf{2) Long-Tailed Risks:} Voice actors face ongoing and evolving risks as their recordings can be continually reused, repurposed, redistributed, and integrated into new AI models long after initial consent, often without their knowledge or further compensation. \textbf{3) Difficulties in Data Traceability and Control:} Voice actors experience a significant loss of control over their voice data post-delivery, with an absence of effective mechanisms to track how their voice files are used, shared, or altered, particularly for AI training or cloning. Based on this context-dependent risk, we propose \textbf{PRAC³} framework, which expands the existing C³ (Consent, Credit,
Compensation) to adapt emerging risks related to voice in Privacy, Reputation, and  Accountability.

%%% Local Variables:
%%% mode: latex
%%% TeX-master: "main"
%%% End:

%  LocalWords:  biometrics cryptographic parallelized lossy
    % basic introduction
\section{Related Work}
\label{sec:relwork}
\subsection{Voice Actors' Risk Beyond the Three Cs} 
Voice actors have become focal points of concern with the rise of AI-generated speech. Unlike text or image data, voice data is biometric, uniquely identifying and intimately tied to an individual. Recent works started to catalog ethical and safety risks posed by synthetic voice cloning from impersonation scams contributing to swatting attacks and financial fraud and in malicious deepfakes that fuel misinformation~\cite{hutiri2024not}. Critically, these risks often arise from multi-facedted interactions among stakeholders e.g. a company releasing a model~\cite{thevergeTikTokSettles}, a malicious actor using it, and victims~\cite{hutiri2024not} and voice owners suffering the consequences, making it hard to assign accountability~\cite{agnew2024sound}. Voice actors face challenges at this intersection of creative labor and biometric vulnerability because their voices are not only artistic output but also their personal identifiers with overlapping risks. Moreover, voice actors are discovering unauthorized voice clones of themselves being distributed online, such as, clone voices of voice actors with foul and offensive language, causing her irreparable harm~\cite{thevergeTikTokSettles} and even celebrities have voiced concerns where Scarlett Johansson's AI-generated voice without consent~\cite{nbcnewsScarlettJohansson}. 

Recently, IAPP advocates stronger protections than those afforded by existing frameworks (3 C's) for visual or textual creators~\cite{iappIAPP}. Notably, when current law often falls short where in the U.S., a person’s voice itself is generally not protected by copyright (only specific recordings are)~\cite{iappIAPP}. For example, simply giving credit to an original artist when an AI mimics their style does not prevent economic displacement or the emotional impact of seeing one’s style copied by a machine~\cite{kyi2025governance}. Likewise, one-time compensation (such as paying for a data license) might not be adequate if the AI model continues to derive value from an artist’s work indefinitely a concern related to the long-tailed nature of generative AI benefits~\cite{sfpublicpressCaliforniaCreatives}. Thus, voice actors are part of a broader creative workforce that encounters multifaceted risks in various forms. A primary client may repurpose voice data beyond the scope of the original agreement, such as for future AI training or product expansion; AI companies often train models on massive repositories of “publicly available” data such as audiobooks or demo reels; secondary creators may use voice samples as background content, remix them into memes, or manipulate them with AI tools for entertainment without authorization~\cite{thevergeTikTokSettles}; adversarial actors have used cloned voices to perpetrate fraud, financial scams~\cite{bateman2022deepfakes, jasserand2024deceptive}.

In our work, we investigate how voice actors face evolving risks beyond consent, credit, and compensation, proposing a dynamic threat model to assess how voice data, once shared, can be continuously reused and repurposed in the generative AI era.
%\subsection{Credit and Compensation in Composition of Privacy in Data Sharing}

\subsection{Privacy Risks in Data Sharing}
Today’s consumer-facing systems, such as recommender systems, search engines, and more recently, large language models, often provide personalized services by processing vast amounts of data. There are two different ways data owner shares their data: deliberately and in an implicit manner. Implicit data sharing happens when users naively consent to terms of service, as is common with applications like ChatGPT~\cite{yu2024exploring}, search engines~\cite{sharma2024m}, and social media~\cite{kaushik2024cross, sharma2023user} platforms, web-crawled public user data~\cite{schuhmann2022laion, baroni2009wacky, thomee2016yfcc100m, baumgartner2020pushshift}. In contrast, deliberate data sharing occurs when individuals knowingly contribute to public datasets~\cite{sharma2023disability, ardila2019common, kearns2014librivox}, often presenting tangible benefits like monetary rewards, research contributions, or social support or as a part of their primary profession~\cite{allen1999privacy, lane2014privacy, board2015sharing, godard2003data}. This includes ML data workers and creative contributors but also participants in open voice data initiatives like Mozilla Common Voice and LibriVox, whose recordings have been foundational to speech technologies~\cite{panayotov2015librispeech, tauberer2010learning}. 
%In some cases, individuals share highly sensitive data, such as medical histories and health records, with public health organizations and nonprofits to support epidemiological research, disease surveillance, risk assessment, and the development of new treatments~\cite{fairchild2007public}.

A major challenge for data contributors is the lack of mechanisms to manage privacy, ownership, and value over time~\cite{sharma2024m, sambasivan2018privacy}. 
In the case of deliberate data sharing, especially for voice, privacy risks are increasingly entangled with issues of long-term control and commodification~\cite{panayotov2015librispeech}. As generative AI models evolve and are reused across multiple applications, original contributors often lose visibility into how their data is being leveraged by whom, and to what end. This shift has led data owners to reframe privacy not only as a matter of control, but also as a function of data valuation and fair compensation\cite{romanosky2009privacy, acquisti2013privacy}.
%Privacy has become more critical for \textit{``deliberate data sharing context''} as more recently privacy is contested with data ownership and valuation overtime with the rise of large language models and their supply chain effect where the model and training data have been purchased and been used by many third parties. 
%On top of that, rapid development of LLM have changed how people perceive privacy, shifting the parameters around privacy risks~\cite{sharma2024m}. As a result, contemporary assessments of risks, along with users' evolving perceptions and behaviors, may differ from traditional views, where people may perceive data sharing as an adversarial lens.  Instead, data owners are increasingly viewing privacy in terms of data valuation or fair compensation~\cite{romanosky2009privacy, acquisti2013privacy}. 

Despite the length of research in data sharing and governance, to date, \textit{there are no solutions that help data owner assess their risk during data-sharing activities}. Existing frameworks such as, such as NIST~\cite{lefkovitz2020nist}, ISO/IEC~\cite{purdy2010iso}, CSA CCM~\cite{p3}, and GDPR~\cite{p5}, though designed for risk assessments of federal information systems and organizations, have not been adapted to provide data owners with tools in understanding risks and the value of their data over time.

%%%%%%%%%%%%%%%%%%%%%%%%%%%%%%%%%%%%%%%%%%%%%%%%%%%%%%%%%%%%%%%%%%%%%%%%%%%%%%%%
\section{Method}
\label{sec:sec_assessment}

We conducted an interview study with 20 voice actors to better understand how generative AI risks, particularly synthetic voice replication, impact voice actors in terms of reputational, privacy and security threats. The study was approved by our university’s Institutional Review Board (IRB).

\textbf{Participants}
Participants were recruited via email, with an explanation of the research and an invitation to participate in an interview. Our sample consisted of freelancers, contract workers, and full-time professionals, with voice acting experience ranging from 4 to 20 years. Each participant was assigned a random number (e.g., P01) for anonymity.

\textbf{Interview Process}
We developed a semi-structured interview protocol that asked participants about their typical workflow and data sharing practices, including how they select platforms for voice work, auditioning processes, contract experiences, and tools used; their experience and awareness of generative AI and the risks related to synthetic voice replication; their opinions on data ownership and consent; and the privacy risks related to their work, such as synthetic voice misuse in unauthorized contexts. 

Where participants raised topics related to our research goal but not covered in our interview protocol, we asked follow-up questions; for example, P4 mentioned experience watermarking voice data, so we asked more about current watermarking practices. Interviews lasted between 40 and 70 minutes and averaged approximately 60 minutes. All interviews were conducted in English. 

\textbf{Data Analysis}
Interviews were recorded and transcribed over Zoom and manually corrected as necessary by the first and second authors. To answer the research questions, we adopted a deductive-inductive approach to coding the interview transcripts. We employed the following deductive codes: Participant Category, Awareness and Understanding of AI Risks, Workflow and Practices, Ownership and Compensation, and Privacy and Security Concerns.

Each interview transcript was coded in several rounds.  The authors first coded seven transcripts and created an initial codebook that included inductively generated themes. After that, the authors reviewed and revised the codebook. For example, the Participant Category code did not exist in the codebook at the beginning, but after the first round of coding, we found that the participants’ experience as voice actors and the resources available to them when encountering legal and data ownership issues varied significantly. As a result, we added the Participant Category code to better capture the diversity of experiences and resources across participants. The authors then coded two additional transcripts at a time until all transcripts had been coded. Finally, the authors followed a thematic analysis process~\cite{clarke2017thematic} to generate themes that answered research questions and created four different user personas, using a shared document to precisely define the themes.

\section{Results}
\subsection{Personas of Voice Actors}
From our list of 20 participants, we created four personas based on years of experience and availability of resources, as shown in Table~\ref{tab:design_preferences}. Four significant personas are: (a) Emerging Professionals (Low experience, Low resources); (b) Solo Defender (High experience, Low resources); (c) Delegator (Low experience, High resources); (d) Strategist (High experience, High resources). For instance, common traits of delegator personas include, more than 5+ years of experience, strong home studio setups, rely on direct client relationships with some occasional platform works, proactive in using AI riders from NAVA and often opt outs if client disagrees. Often they have their own representative to assess contracts and negotiate client compliance with AI clauses. This group hold deep concerns over AI risks, specially unauthorized cloning, unauthorized voice usage for AI training.
P12 (Delegator personas) shared their experience explaining how their agent assess the project 
\begin{quote}
\textit{``You're handed a contract, and you glance at it, and you know you got to sign it there ... I didn't go over it with a fine tooth comb. In one of TV affiliates projects, my agent asked, we need to confirm that you're not going to use this for AI. They said 
we're not going to confirm anything. And my agent and I were both-we're not going to do this''}. 
\end{quote}

Whereas solo defender involves voice actors from mid-level experience to senior freelancer. They have reasonable knowledge of tech, and media IP. They handle their own contract. They have systemic awareness of AI risk across sectors. They share common pitfalls, silent data scraping from public platforms and lack of enforceable global standards for dataset auditing. P5 said
\begin{quote}
\textit{``I did get an offer, and I try to research it as much as possible. And I do ask, you know, what is this going to be used for... before I engage in even auditioning.''}
\end{quote}

\begin{table}[ht]
\caption{Personas of Voice Actors}
\label{tab:design_preferences}
\begin{tblr}{
  colspec={lX},
  hlines,
  row{1} = {font=\bfseries},
  row{2,4} = {bg=gray!10}, % Optional row shading for readability
}
Participant Category & Categorization based on resources (high/low) and experience (high/low) \\
Solo Defender & High experience, Low resources. Difficulty identifying risky clients, hidden scraping, no enforcement power. \\
Emerging Pro & Low experience, Low resources. Accept unclear terms, fear AI misuse, lack AI literacy. \\
Strategist & High experience, High resources. Concerns about AI misuse affecting personal brand; proactively negotiate contracts. \\
Delegator & Low experience, High resources. Relies on agents/platforms, vague understanding of AI risks. \\
\end{tblr}
\end{table}

\subsection{Indication of Risk Through Interaction}
\label{sec:RQ1}
\begin{figure*}[htbp]
    \centering
    %\framebox[\linewidth]{For Measuring!}
    \includegraphics[width=\linewidth]{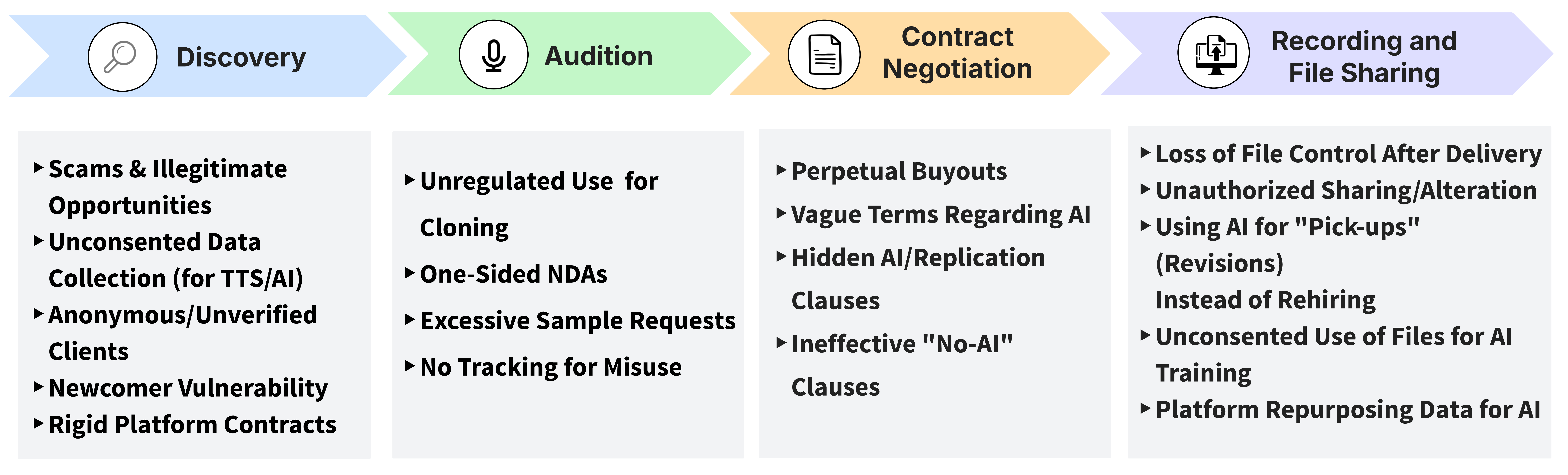}
    \caption{Risks and AI-related threats in different stages of voice acting work, including discovery, audition, contracting, recording and file sharing. 
    %\tanusree{make the caption self explanatory. what it represent }
    }
    \label{risk-phase}
\end{figure*}

In their routine professional activities, voice actors work across a range of sectors such as, commercials \& advertising, followed by audiobooks, animation \& cartoons, and E-Learning \& educational content, video games, and podcasting \& audio dramas, dubbing \& localization, live performance \& theatrical productions, each showing high demand across the industry. We found a reservation of voice actors to work on Text-to-Speech (TTS) \& AI Voice Training project due to both normative and practical concerns. For all participants, voice acting was their primary profession, typically following a structured workflow: 
%\yihao{this is not aligned with the figure and the following subsection division.}
(a) discovery of the work; (b) Audition ; (c) Contracting; (d) Recording and File sharing. In this section, we lay out the risks in each stage of their interaction, in particular risks pertaining to with advanced AI landscape, as shown in Figure \ref{risk-phase}. 

\textbf{Discovery.} The first phase involves voice actors seeking projects aligned with their skills, interests, and availability. Participants identified three primary channels for finding work: commercial platforms (e.g., Voice123, Amazon ACX), social media job postings (e.g., Facebook, Twitter, LinkedIn, Bluesky, Discord), and agent representation.

%\underline{Risk}
In this phase, participants explained a key concern, which was the difficulty in distinguishing legitimate opportunities, particularly when platforms allow anonymous clients. Thus, determining if the project is legit or a means to collect voice data for unconsented use in Text-to-Speech (TTS) applications, often challenging for voice actors. %On top of that, only relying on standardized platform contracts further restricts negotiation, leaving them with the binary response of either accept the terms or forgo the work. 
Participants expressed a preference for working with agents or clients who engage in direct, identifiable communication, allowing for dialogue and verification in contrast.  However, a very small number of participants had access and the means for agents or direct connections with publishing houses and individual clients, which largely contingent on the actor’s experience and the professional network they had developed over time. This highlights a level of gatekeeping and structural advantage not available to less experienced voice actors, leading to risks of their voiceprint ending with bad actors. 
%An ad-hoc approach they take is to look for a number of completed projects in the client profile. 

\textbf{Audition.} Auditions serve as a gateway for voice actors to secure roles, yet this crucial stage remains largely unregulated in terms of how submitted samples may be used beyond the selection process. Participants mentioned producers or client often shielded by non-disclosure agreements (NDAs) signed by voice actors, where the actors themselves typically operate without reciprocal legal protection. The current industry norm relies on informal trust. 
%As P18 said-
%\begin{quote}
    %\textit{``No, you have to sort of trust that they're going to treat it as an audition. No one demanded anything along those lines [re: AI use during audition]. It doesn’t factor in.''}
%\end{quote}
We encountered incident where audition sample were indeed used, but later remediate by booking the artist, as P11 recounted, 
\begin{quote}
   \textit{``It’s just an unspoken agreement, you have to sort of trust... I’ve only had once-client used my audition for the job. But they booked me later, so it was fine.''}
\end{quote}
Participants consistently identified this auditioning phase as one of the most vulnerable to misuse particularly regarding unauthorized AI training or voice cloning.
To mitigate these risks, voice actors rely on informal, ad-hoc strategies, such as, mark as read flag if (a) requests for unusually long audition samples (b) lack of communication from intermediaries (e.g, agent, client, casting director). As P18 said 
\begin{quote}
   %\textit{``Usually, there’s one or two middle people… I send my demo to a agent and age to casting person, who sends it to the client. when I see abrupt stop of communication, i get the signal of misuse.''}
   \textit{``Usually when you do a read like that, it's between a minute to 5 minutes long based on the project types like, commercial might need longer one where audiobook should not need more than 60s. If its longer, there is something off with the client.''}
\end{quote}
Some participants mentioned technical deterrents, such as inserting beeps into their samples to prevent unauthorized use. However, most rejected this approach due to concerns that it compromised the quality and may jeopardize job opportunities. Notably, several participants suspected that their auditions had indeed been misused, yet felt powerless due to no means of tracking or legal recourse.

\textbf{Contract Negotiation.}
In contract negotiations, voice work involves credit, consent, and compensation. Participants reported credit practices, such as,  audiobook clients often provide public-facing credit to narrators, but rare in commercial or corporate contexts. They mentioned varied compensation including, session-based fees, time-limited buyouts, and perpetual rights. In the era of generative AI, compensation terms like \textit{``in perpetuity buyout''} are seen as red flags, often signaling potential for AI repurposing if pay is not high. P12, the original male voice of a voice assistant tool of major tech, recounted how a one-time session and a yearly non-compete fee evolved into widespread, unauthorized use of his voice. 
\begin{quote}
    \textit{``
    It was released 5 or 6 years back. I regret not having a lawyer review the contract, which included broad terms like 'in any form or technology now known or unknown, in perpetuity.' I later found out my voice is rented to Y and Z companies. Interestingly got to know that from my daughter and friend. That didn’t feel good. I hadn’t understood how my voice would be used, but for a while, people kept asking if I did a hotel ad in Berlin or other projects. My voice ended up in explainer videos, commercials, and even a video game chatbox without additional pay. Since then, I’ve renegotiated for higher compensation. Still, the original deal locked me into a much lower rate especially compared to the female voice, who reportedly earns around \$250k a year.''}
    
   % They pay me a do not compete fee every year. means that I not be a digital assistant for another high profile product. I wish I had had a lawyer look over the contract years back. They used language like in any form or any technology now known or unknown in perpetuity. I found out my voice is rented out to speechify, interestingly from a family member which was not a good feelings. The contract that I signed with them. I did not. I didn't know how this will be use. But for a while there I had people sending me emails or texts about once or once a week and say, did you do a commercial for a hotel in Berlin. I ended up on lots of explainer videos. And actually a friend of mine and his son play video games. There in the chatbox, automatic reader, was my voice. So all these applications all over the place, and I was not being paid for those. After that, What I've done is negotiate up the rate that they pay me. The frustrating thing is that the person who negotiated with me originally kept me at what I didn't know was an extremely low level where the female voice get around \$250k a year, mine is lot lower even after this mishandling and negotiation.''}
\end{quote}
This highlights mismatch between contractual terms and long-term value in the AI era. 
Participants are becoming more vigilant about ownership risks. As P3 noted, actors often refer to \textit{``voice ownership''} without fully understanding its implications.
\begin{quote}
    \textit{``We have signed away the right to the performance and not the voice print.''}
\end{quote}

%As most of the participants including P1 and P6 said -
% \begin{quote}
%     \textit{``Some contracts now say, ‘We will not use your voice for AI’.. or I sent the AI rider that says none of the work will be used for generative AI recreation.''}
% \end{quote}
To address concerns with three Cs, participants cited the NAVA community-developed AI rider, which is becoming a standard to attach during contracts. Several participants shared troubling experiences. P1 described a case where their voice with client for a video game was later synthesized with AI despite a \textit{``No-AI''} clause the developers never saw. P4 also recalled a contract in which a clause allowing voice replication was buried in Exhibit A, unseen by the actors. Such cases highlight how vague or inaccessible legal language leaves voice actors, especially newcomers vulnerable to exploitation.
\begin{quote}
    \textit{``Exhibit A wasn’t even seen by the actor. The publisher signs it''}
\end{quote}

\begin{figure*}[htbp]
    \centering
    %\framebox[\linewidth]{For Measuring!}
    \includegraphics[width=\linewidth]{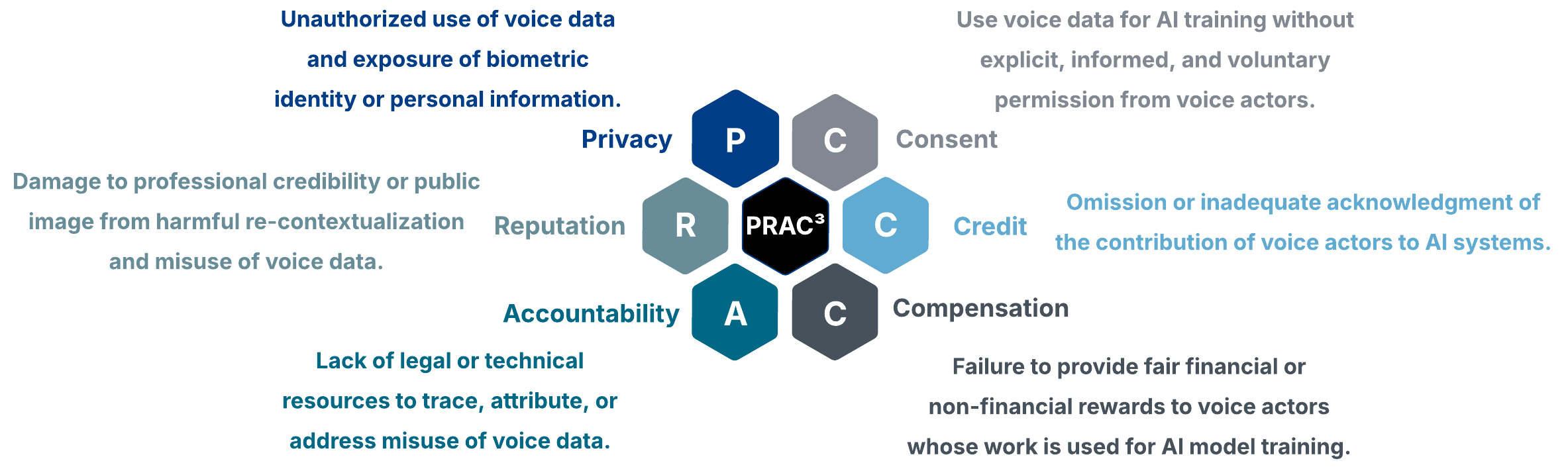}
    \caption{The PRAC³ framework, including six risk dimensions in the use of voice actors' data in the context of generative AI: Privacy, Reputation, Accountability, Consent, Compensation, and Credit.
    %\tanusree{make caption self explanatory. read section "Conceptualized Framework of Threat Model to Assess Risks" and understand what it means by accountability.}
}
    \label{PRAC3}
\end{figure*}

\textbf{Recording \& File Sharing.}
In this phase, a common set up of voice actors is - recording in home studios using XLR mics, audio interfaces, and DAWs like Reaper or Audacity. They typically record in three model (a) home studio, (b) remote sessions via Source Connect, or (c) on-site studio sessions. They share files through platforms like WeTransfer (valued for simplicity and notifications), Dropbox, Google Drive (for large or ongoing work), or email (for small .mp3 auditions). They mentioned platforms like Voices.com or ACX sometimes handle the upload internally. This stage carries significant risks, primarily loss of control and transparency. Once files are sent, actors have little visibility into how they are used, shared, or altered. As P14 noted:

\begin{quote}
    \textit{``There’s no way to verify that a client sticks to their usage agreement... unless I catch it in the wild. Once it’s downloaded... they might share with someone else, chop it up, and send it off. I don’t think none are using such technologies in industry now. Maybe there could be something in that initial audio file... so that if an AI clone is made, it’s detectable.''}
\end{quote}
Participants also raised concerns about "pick-ups"—where clients use AI to mimic voices for minor revisions rather than hiring the actor again. Many expressed frustration that no watermarking or traceability measures are in place to detect such misuse. Further platform-specific concerns emerged, for example, P8 highlighted potential conflicts of interest with Amazon ACX, questioning whether submitted voice data could be repurposed for amazon's product. Across the board, actors emphasized the lack of post-delivery tracking and growing fear of their recordings being used for AI training.

%\tanusree{need a professional diagram for these phase with risks involved}
%\textbf{Post-Delivery.}

\subsection{Current \& Long Term Risks Perception} 
Our analysis with professional voice actors revealed awareness of both current vulnerabilities and future threats, beyond the three Cs, particularly around privacy, reputation, and accountability.

\textbf{Security and Identity Concerns}
 Participants expressed growing concern over the biometric nature of voice data and the ease with which it can now be cloned and reused without authorization. Several actors identified the potential for fraud and impersonation particularly, in financial or emergency contexts. For instance, P16 noted 
 \begin{quote}
     \textit{``Scammers can now... call you and say ‘Mommy, I’m being hurt’ using your kid’s voice. And you don’t know if it’s real. Its really frightening. My voice is out there more than an average users.''}
 \end{quote}
Some reported concerns on voice authentication in banking. Meanwhile, actors like P16 pointed to the existential challenge of deepfakes, describing it as \textit{``Not being able to verify your own voice because someone has stolen it.. next-level voice theft.''}. P8 explained - 
\begin{quote}
    \textit{``If financial institutions use voices... that’s not a good idea considering how easy it is to duplicate. I also sometime wonder- banks that ask for voice verification... Is it being used to train something else?''}
\end{quote}
One participant with a cybersecurity background (P6) emphasized that some deepfake uses cross into serious crime, noting incidents where AI-cloned voices were used for \textit{``swatting''} (calling in fake threats) and other dangerous hoaxes. 
These concerns underscore the shift from theoretical risk to practical harm, particularly for security and safety of voice actors in their personal life. 

%where many stressed that voice-cloning attacks are no longer hypothetical, they \textit{``have happened and continue to happen.''}

\textbf{Reputational and Ethical Risks}
Voice actors also raised serious concerns about their voices being used in ways that contradict their values and can often damage their personal standing. We found scenarios where some participants found their voice being mismassed to create AI-generated voice content in controversial media such as, political,  controversial media. One actor recalled a case where P4 mentioned-
\begin{quote}
    \textit{`` I initially worked on a anime character which was normal. then they made that character do AI-generated porn... that reflects badly on me, which was never consented.''}
\end{quote}
Some also feared their voices could be embedded in propaganda or defamatory content, with no clear mechanism for recourse or correction. P17 described an unsettling experience of hearing accidentally a TV commercial on political agenda in gender issues which sounded like her own voice which she never recorded. 
% \begin{quote}
%     \textit{`` I was sitting in my couch and suddenly recognizing my own voice from a TV commercial, speaking phrases I never recorded. the voice was mine, the words were not.''}
% \end{quote}
This lack of control over one’s digital likeness raises questions about the professional and personal boundaries in the age of generative AI.

\textbf{Accountability and Legal Uncertainty.}
%The legal ambiguity surrounding voice cloning and synthetic content emerged as a recurring theme a.
Participants expressed frustration over the lack of enforceable rights and mechanisms to trace, remove, or contest the misuse of their voice. For example, P117 described a situation in which a TikTok user initially perceived as a fan used a voice sample from her website to create a reel video:
\begin{quote}
    \textit{``At first, I hear my voice in the background, it seemed benign. Then I realized there was AI to clone certain words I never said. If those memes become more extreme, who is accountable-- me, the person who cloned, or the TikToker?'}
\end{quote}

Beyond the concerns of accountability, some participants added concern of professional and economical reputation. P17 highlighted how their voice association with low-quality productions or cloned by individuals could damage his credibility, as audiences might conflate the synthetic performance with the original artist. Similarly, P3 explained 
%the opacity of content distribution chains and the inadequacy of existing legal measures:

\renewcommand\arraystretch{1.2} 

\onecolumn
\begin{center}
\small
\begin{longtable}{@{}c m{3.5cm} m{4.0cm} >{\raggedright\arraybackslash}m{8cm}@{}}
  \caption{Examples of Reported data-misuse and AI-related incidents affecting professional voice actors.}
  \label{demographic}\\
  \toprule
  \textbf{ID} & \textbf{Scenario} & \textbf{Incident (Participant)} & \textbf{Analysis using PRAC³ Framework} \\
  \midrule
  \endfirsthead

  \multicolumn{4}{l}{\textit{(Continued from previous page)}}\\
  \toprule
  \textbf{ID} & \textbf{Scenario} & \textbf{Incident (Participant)} & \textbf{Details} \\
  \midrule
  \endhead

  \midrule \multicolumn{4}{r}{\textit{Continued on next page}}\\
  \endfoot

  \bottomrule
  \endlastfoot

  1 & Audition sample reused in national commercial
    & P17 discovered her voice in an ad she never recorded (P17)
    & \textbf{PRAC³ Domain:} Consent, Compensation, Accountability \newline
      \textbf{Threat Agent:} Client/Studio \newline
      \textbf{Asset at Risk:} Voice data, creative labor \newline
      \textbf{Potential Impact:} Unauthorized commercial use; loss of income; reputational risk \newline
      \textbf{Mitigation Status:} None – discovered post-facto \\
  \midrule

  2 & Voice used in AI-generated adult content
    & Game mod used AI to create pornographic scenes with actor's voice (P7)
    & \textbf{PRAC³ Domain:} Reputation, Consent, Accountability \newline
      \textbf{Threat Agent:} Third-party modders \newline
      \textbf{Asset at Risk:} Public persona, moral integrity \newline
      \textbf{Potential Impact:} Defamation; emotional distress \newline
      \textbf{Mitigation Status:} Unreported; no recourse \\
  \midrule

  3 & Exhibit A clause allows post-production cloning
    & Audiobook contract allowed voice replication without notice (P4)
    & \textbf{PRAC³ Domain:} Consent, Compensation, Accountability \newline
      \textbf{Threat Agent:} Publisher \newline
      \textbf{Asset at Risk:} Voice likeness; residual earnings \newline
      \textbf{Potential Impact:} Job displacement; IP erosion \newline
      \textbf{Mitigation Status:} Discovered post-signing \\
  \midrule

  4 & AI voice scam using child’s cloned voice
    & Scam calls using cloned voice of loved one (P16)
    & \textbf{PRAC³ Domain:} Privacy, Identity, Accountability \newline
      \textbf{Threat Agent:} Cybercriminals \newline
      \textbf{Asset at Risk:} Biometric identity \newline
      \textbf{Potential Impact:} Financial fraud; emotional harm \newline
      \textbf{Mitigation Status:} Hypothetical/precautionary \\
  \midrule

  5 & Podcast platform AI-translates and clones voice
    & Large tech [Y]  translated podcaster’s voice without opt-out (P19)
    & \textbf{PRAC³ Domain:} Consent, Privacy, Accountability \newline
      \textbf{Threat Agent:} Platform provider \newline
      \textbf{Asset at Risk:} Voice data; linguistic identity \newline
      \textbf{Potential Impact:} Unconsented speech generation \newline
      \textbf{Mitigation Status:} Actor manually obstructed usage \\
  \midrule

  6 & No disclosure of voice reuse for AI training
    & P4 reported clause only found post-distribution
    & \textbf{PRAC³ Domain:} Consent, Privacy, Compensation \newline
      \textbf{Threat Agent:} Client \newline
      \textbf{Asset at Risk:} Voice training data \newline
      \textbf{Potential Impact:} Unpaid AI training use \newline
      \textbf{Mitigation Status:} No consent captured \\
  \midrule

  7 & AI-generated voice used in foreign language translation
    & Large tech [Y] used AI to translate podcaster's voice without clear opt-in (P16)
    & \textbf{PRAC³ Domain:} Privacy, Consent, Accountability \newline
      \textbf{Threat Agent:} Platform \newline
      \textbf{Asset at Risk:} Voice identity; language authenticity \newline
      \textbf{Potential Impact:} Loss of control over voice use, misrepresentation \newline
      \textbf{Mitigation Status:} Voice actor manually obstructed feature with background audio \\
  \midrule

  8 & Audition samples used without hiring actor
    & Actors heard their audition voices in released work (P14, P16, P17, P18, P20)
    & \textbf{PRAC³ Domain:} Consent, Compensation, Credit \newline
      \textbf{Threat Agent:} Client/Producer \newline
      \textbf{Asset at Risk:} Audition recordings; performance data \newline
      \textbf{Potential Impact:} Unpaid labor; reputational confusion \newline
      \textbf{Mitigation Status:} Typically undiscovered until after release \\
  \midrule

  9 & Voice used in modded game porn content
    & AI-generated adult content using voice actors' characters (P7)
    & \textbf{PRAC³ Domain:} Reputation, Privacy, Accountability \newline
      \textbf{Threat Agent:} Third-party users \newline
      \textbf{Asset at Risk:} Character alignment; public image \newline
      \textbf{Potential Impact:} Moral distress; brand damage \newline
      \textbf{Mitigation Status:} No action taken; actors unaware until fans reported \\
  \midrule

  10 & Hidden AI training clauses in audiobook contracts
    & Exhibit A allowed voice replication post-recording (P4)
    & \textbf{PRAC³ Domain:} Consent, Accountability, Compensation \newline
      \textbf{Threat Agent:} Publisher \newline
      \textbf{Asset at Risk:} Creative control; residuals \newline
      \textbf{Potential Impact:} Job replacement by AI; under-compensation \newline
      \textbf{Mitigation Status:} Clause discovered only post-facto \\
  \midrule

  11 & Client reuses voice clip across projects without permission
    & P17’s voice reused in ad without consent
    & \textbf{PRAC³ Domain:} Consent, Accountability, Credit \newline
      \textbf{Threat Agent:} Client \newline
      \textbf{Asset at Risk:} Vocal performance; authorship \newline
      \textbf{Potential Impact:} Unauthorized branding; reputational risk \newline
      \textbf{Mitigation Status:} No prior notification; discovered incidentally \\
  \midrule

  12 & Scam calls using AI voice cloning of relatives
    & Actors fear scammers using their voice for fraud (P3, P16)
    & \textbf{PRAC³ Domain:} Privacy, Identity, Accountability \newline
      \textbf{Threat Agent:} Cybercriminals \newline
      \textbf{Asset at Risk:} Biometric voice identity \newline
      \textbf{Potential Impact:} Financial scams; family trauma \newline
      \textbf{Mitigation Status:} No technical prevention mechanisms \\
  \midrule

  13 & AI contracts lack explicit voice usage limitations
    & Contracts omit AI voice use clauses (P14, P1)
    & \textbf{PRAC³ Domain:} Consent, Privacy, Accountability \newline
      \textbf{Threat Agent:} Clients/Platforms \newline
      \textbf{Asset at Risk:} Legal rights over voice data \newline
      \textbf{Potential Impact:} Non-consensual reuse or AI training \newline
      \textbf{Mitigation Status:} Actors often overlook contract language \\
  \midrule

  14 & Perpetual license buried in email agreements
    & Clients assume full rights from email threads (P10, P18)
    & \textbf{PRAC³ Domain:} Consent, Compensation, Credit \newline
      \textbf{Threat Agent:} Clients \newline
      \textbf{Asset at Risk:} Work ownership; royalties \newline
      \textbf{Potential Impact:} Lack of residuals; misappropriation \newline
      \textbf{Mitigation Status:} No formal legal review of communication \\
  \midrule

  15 & Replacement by AI for minor roles or demo work
    & Lost work for minor roles to AI-generated voices (P14)
    & \textbf{PRAC³ Domain:} Compensation, Reputation, Accountability \newline
      \textbf{Threat Agent:} Clients \newline
      \textbf{Asset at Risk:} Job opportunities; creative career pathways \newline
      \textbf{Potential Impact:} Job displacement \newline
      \textbf{Mitigation Status:} Community advocacy; union action (no technical protection) \\
  \midrule

  16 & Voice licensed and mass redistributed via third-party
    & Large tech company licensed actor's voice to third-party platforms (P12)
    & \textbf{PRAC³ Domain:} Consent, Compensation, Accountability, Privacy \newline
      \textbf{Threat Agent:} Clients \newline
      \textbf{Asset at Risk:} Voice data; public image \newline
      \textbf{Potential Impact:} Ongoing uncompensated use; loss of control; reputational risk \newline
      \textbf{Mitigation Status:} Attempted renegotiation failed \\
\label{tab2}
\end{longtable}
\end{center}
\twocolumn
\begin{quote}
    \textit{``I don’t doubt one day some content’s gonna feature my voice … and I’m very much scared for that day to navigate legal world... more scared when legitimate companies and criminals alike, now a temptation to “rip off everybody” by harvesting voices, and our legal system is only starting to grapple with it.''}
\end{quote}
These difficulties were particularly severe for non-union actors, who frequently did not have the financial or institutional backing necessary to explore abuses or seek redress. 
With many intermediaries, such as, casting agents, platforms, production studios for a project, standing between them, identifying source of harms, and tracing accountability becomes a near-impossible task.

\subsection{Conceptualized Framework of Threat Model to Assess Risks} 
%\tanusree{need a professional diagram for the new framework within threat modeling context for future risk assessment}
%\textbf{Motivating Case}. Below is a case to illustrate the threat model on data sharing 
As the voice industry intersects increasingly with generative AI, voice actors face distinct and compound risks to their identity, labor, and safety. These risks are structural, embedded in how digital labor is extracted, synthesized, and monetized. 

Based on the earlier sections on risk indicators through different phases of voice actors interaction to digital platforms as well as their experienced and perceived risks, we proposed a PRAC³ framework. This offer a conceptual tool for threat modeling these long-tailed risks, especially in assessing harms that emerge over time and beyond contractual boundaries. PRAC³ stands for \textbf{\textit{``Privacy, Reputation, Accountability, Consent, Credit, Compensation''}}. Each dimension represents a critical vector of exposure or harm for voice actors in the AI data economy. \textbf{Consent, Credit, Compensation} presents foundational rights which often overlooked or bypassed in AI data pipelines. newly added components from voice actor's experience: \textbf{Privacy} which presents breaches of biometric identity through cloning or surveillance; \textbf{Reputation}, which represents harm from voice misuse in misaligned, offensive, or deceptive contexts and finally; \textbf{Accountability} which present legal and technical gaps in traceability and recourse when voice actors data is misused by adversarial actors and harm general users.

Voice actors experience \textbf{three archetypal threat scenarios} that encapsulate both direct and downstream risks. These scenarios highlight how harm is not limited to the moment of data creation but often arises through redistribution, secondary use, and platform-driven commodification.

\textbf{(a)Voluntary, non-monetary contributors:} Actors donate voice data for public good, only to have it later surface in unauthorized commercial tools.

\textbf{(b) Monetized contractual contributors:} Initial legal agreements include ambiguous language often enabling resale, transfer, or indefinite reuse of voice data, especially following corporate changes.

\textbf{(c) Secondary, informal misuse:} Legally recorded voices leak into meme culture, satire, or political propaganda via AI tools, distorting public perception and damaging actors’ professional standing.

Across all scenarios, key assets are voice recordings with identifiable voice features, voiceprint which is a unique vocal fingerprint capable of identification or cloning, reputational credibility, and contractual protections. When a voice actor performs, they manipulate multiple acoustic and articulatory signals to create different characters, emotions, or identities. These changes affect the perceived voice, but the underlying biometric voice signature often remains partially detectable by machines (e.g., AI voice recognition); even if data is anonymized before being shared, advanced analytics or cross-referencing with other datasets could re-identify contributors. 

%In the above cases, stakeholders may misuse the data, or hackers targeting the dataset could steal or corrupt the data, either while it is stored in databases or transmitted between platforms, selling it to third parties without explicit consent from contributors. The major entry point of threat contributors might unknowingly expose sensitive details during the usage of the service or sharing data; even if data is anonymized before being shared, advanced analytics or cross-referencing with other datasets could re-identify contributors. 

Table~\ref{tab2} illustrates PRAC³ framework by mapping real-world incidents shared by voice actors to the six dimensions. Each case illustrates how risks unfold across time and contexts: P7's incident where a modder used AI to generate explicit content using a recognizable voice from a game voice character a violation of Reputation, Consent, and Privacy.
%\textbf{Privacy.} Privacy concerns focus on the control voice actors have over the collection, storage, and downstream use of their biometric voice data. Unlike passwords or email addresses, voice is a non-revocable identifier that can be easily harvested from public media, demos, or auditions and used to train synthetic voice models without explicit consent. This risk can be amplified for voice actors, who often have large quantities of high-quality speech data publicly accessible, increasing their exposure far beyond that of the average digital user.
%\tanusree{add a table}

%%%%%%%%%%%%%%%%%%%%%%%%%%%%%%%%%%%%%%%%%%%%%%%%%%%%%%%%%%%%%%%%%%%%%%%%%%%%%%%%
%\section{Tool Development}
%\label{sec:tool_dev}

%\section{Results}
%\label{sec:results}
\section{Discussion}
\label{sec:conclusion}
%%%What is the big take away from your research. Include any limitations or future work here. 

\subsection{Ethical Frameworks: From C³ to PRAC³} 
Our work broadens the discussion of ethical AI data use by expanding the \textit{``C³''}(Consent, Credit, Compensation) to PRAC³, adding Privacy, Reputation, and Accountability, which emerged through our findings and are important dimensions for long-term risk assessment. Prior work centered on creators' consent to their data and receive attribution and payment~\cite{kyi2025governance, blaising2022managing}. PRAC³ model can capture context-transcending risks posed by generative AI, for instance, how voice actors’ \textit{``vocal identities''} can become decoupled from context, authorship, and control in AI systems.
Our findings reveal that voice, as a unique identifier, can be misused by clients or downstream users, causing harm to contributors’ personal and professional identities. PRAC³ thus reframes voice actors as stakeholders, not just content sources, to offer a comprehensive model for assessing risks.

Privacy, as a pillar, encourage rethinking voice data not merely as creative output but as biometric personal data. Voiceprints which is central to voice actors’ identity, are often scraped or shared without consent, echoing Zuboff’s "surveillance capitalism," where human experience becomes unconsented raw material~\cite{zuboff2023age}.
Our findings present that voice actors' sign a contract for their voice performance, not the voiceprint. Despite growing legal recognition (e.g., Illinois’ BIPA~\cite{cook2024illinois}, EU AI Act~\cite{europaActFirst}, CCPA~\cite{CCPA}), our findings reveal widespread misuse, particularly in privacy, security, and safety, due to a lack of provenance. Once voice data is embedded in models and spread across platforms, it's nearly impossible to trace or retract. Unlike image watermarking, to the best of our knowledge, robust voice provenance tools remain undeveloped~\cite{pantiukhov2024enhanced, kang2022zk}. Legal protections lag, with gaps illustrated by the TikTok text-to-speech case, where a voice actor’s work was repurposed without her knowledge~\cite{thevergeTikTokSettles}. 
Further, we found voice data reused in controversial memes, raising unresolved questions of accountability regarding whether to attribute the harm to secondary content creators who used the voice sample or the original voice actors whose voice been used. This indicated reciprocal reputational harm for voice actors. By positioning voice data as personal data tied to privacy, reputation, and accountability, our work advances frameworks for voice data governance in AI.

\subsection{Long-Tailed Risks and Novel Threat Modeling}
A key contribution of this paper is its threat modeling of long-tailed risks that emerge over time and across institutional boundaries, beyond immediate consent violations~\cite{cultureandcodeCx2019sVoice}, by highlighting the need for anticipatory risk assessment, mainly explored in high-stakes security areas~\cite{darpaA3MLAnticipatory, repecAnticipatoryGovernance}. PRAC³ framework advances this by integrating baseline risks (C³: Consent, Credit, Compensation) with evolving and future threats with a context-dependent manner~\cite{nissenbaum2004privacy} where each risk scenario defines assets (e.g., voiceprints), identifies threat actors, system or human vulnerabilities, and potential consequences for voice actors such as identity misuse, reputational damage, and clarify assumption and boundary for threat models. 
Our qualitative data illustrates these risks, where a voice actor's character was used in AI-generated pornography; others found their voices endorsing political messages or were rented internationally without consent. Such decontextualized deployments lifted voices from their original intent and violate personal and professional credibility. PRAC³ addresses these temporal and cross-context risks of how privacy is compromised through non consensual AI training, reputation is damaged by misaligned use, and perpetuates fraud and scam.

In effect, PRAC³ functions as a forward-looking framework to encourage 
Data ethics beyond bias and fairness audits, toward systemic. 
We believe this framework will provide means for researchers and practitioners to anticipate low probability but high impact events and institutional failure points (such as when voice data travels through many hands and jurisdictions), rather than leaving voice actors to fight case-by-case battles. 

\subsection{Digital Labor, Exploitation and Precarity}
Credit and Compensation in PRAC³ affirms that voice data constitutes creative labor, situating voice acting within broader critiques of digital labor and platform exploitation. Our interviews reveal systemic issues familiar in gig economies, such as,  power asymmetries, opaque contracts, and precarious work~\cite{zhang2022algorithmic, liang2024valuation}. Many voice actors sign away rights \textit{in perpetuity} for a one-time fee, often without understanding long-term implications with a lack of union protection. Unlike typical gig workers, voice actors go through auditions, yet legal safeguards are often absent in this stage. While AI systems can now learn from minimal samples, our study suggested this phase can lead to risk in displacing entry-level jobs, such as audiobook narration, with low-emotion synthetic alternatives and potentially eroding future labor opportunities, especially for newcomers.
 %Many voice actors operate as freelancers or through platforms, often without union protection, which leaves them vulnerable to unfair terms such as contracts demanding rights “in perpetuity” for a one-time fee.However, unlike many gig workers, voice actors have an auditioning process which doesn't involve any legal or contracting measure leave voice actors vulnerable as advances AI tool can learn from a sample and sever purpose for low level tasks like audiobook which might not need the level of emotion of a human narrator, therefore risking the future labor opportunities for voice actors, particularly who are new and emerging and start with small or basic projects like audiobook narration. 

Our findings show voice actors frequently unaware of how their recordings would be reused. Some voluntarily contributed to early datasets (e.g., LibriSpeech~\cite{kearns2014librivox}), which were later pivotal in training generative AI.  Yet, a decade later, these same contributions have exposed voice actors to a range of risks. In effect, voice actors’ labor is being commodified and endlessly monetized by others, a pattern of algorithmic exploitation comparable to how other AI training data (art, code, writing) have been harvested without rewarding the original creators. Another reflection from our study, \textit{``I wish I had a lawyer 5 years back''}, referencing how their voice was later “rented out” in global ads and games, without compensation. This underscores a critical mismatch between initial contract terms and the enduring value of data in AI systems.
This exemplifies a mismatch between contractual terms and long-term value in the AI era. As Gray and Suri (2019) and others have noted, AI systems are fueled by millions of underpaid workers performing repetitive tasks under precarious conditions~\cite{gray2019ghost, sharma2025politics, sharma2024inclusive}. 

%Our study also presented instances where voice actors unaware of the future use of their data, regardless voluntarily or paid options. For instance, a unique case where voice actors with their free will donated their voice in early large-scale audio datasets, such as LibriSpeech on 2015, later instrumental in the development of speech technologies, including audiobooks and voice assistants and now been used for Generative AI training for many purposed they never thought of. Yet, a decade later, these same contributions have exposed voice actors to a range of risks one of which is loss of employment in longer term. In effect, voice actors’ labor is being commodified and endlessly monetized by others, a pattern of algorithmic exploitation comparable to how other AI training data (art, code, writing) have been harvested without rewarding the original creators. Another common incident from our work is when voice actors say these word \textit{``I wish I had a lawyer 5 years back''} hinting to incident like the derivative of their voice by AI “rented out” to numerous third parties and appeared in ads and video games worldwide, none of which they were paid for.This exemplifies mismatch between contractual terms and long-term value in the AI era. As Gray and Suri (2019) and others have noted, AI systems are fueled by “millions of underpaid workers…performing repetitive tasks under precarious conditions~\cite{key}.
PRAC³ expands the ethical lens by connecting credit and compensation with reputation and labor precarity. Voice actors fear that low-quality synthetic or cheaply made AI clone reproductions of their voice can erode the actor’s credibility, adding a new layer of workplace harm, reputation damage, and market displacement, atop the more traditional concerns of missing credit or payment. Currently, SAG-AFTRA and IATSE have pushed for contractual protections ensuring consent, credit, and compensation for performers’ contributions to AI systems~\cite{iatseIATSEWelcomes}. PRAC³ framework provides a conceptual backbone for these demands, factoring in three new component reputation, accountability, and privacy.

%\subsection{Future Directions and Interdisciplinary Research}
%We envision future research at the intersection of technology, law, and ethics to operationalize the PRAC³ framework. Technologists can work on solutions for provenance tracking and content authentication, for instance, developing algorithms to detect a specific voice fingerprint in generated audio or watermarking techniques that embed source identifiers into voice models without degrading quality with “traceability” and “enforceable boundaries”. Legal scholars and policymakers, on the other hand, can explore new regulatory instruments , such as, concrete classification of “in perpetuity” AI usage clauses and comparative policy analysis – e.g., studying how data protection laws (GDPR, BIPA, etc.) might be extended or interpreted to cover voice AI scenarios.  Ethicists and social scientists could build on this work by investigating long-tailed AI risks in other domains beyond voice. In our immediate next step, we will engage stakeholders in co-designing AI governance mechanisms to operationalize participatory risk framework PRAC³ that pure top-down regulation might miss.
%%% Local Variables:
%%% mode: latex
%%% TeX-master: "main"
%%% End:

\section{Conclusion}
We shed lights on an overlooked area of creative work related to voice which is the current and upcoming commodities in Generative AI model advancement. By interviewing voice actors, we uncover risks of how voice data functions as both a creative product and a biometric marker, placing this community for prolonged challenges that go beyond conventional consent, credit, and compensation. We introduce PRAC³ framework to expand the ethical landscape to include Privacy, Reputation, and Accountability to offer a holistic model for assessing and mitigating these risks. As AI technologies continue to evolve and commodify human expression, this work underscores the urgent need for governance models, legal protections, and technical solutions that center the rights, identities, and labor of voice actors. Tackling these issues is crucial not only for their personal and professional dignity but also for establishing a just and reliable synthetic media landscape.
%\newpage
\section{Ethics Statement}
The PRAC³ framework proposed in this paper is derived mainly from the information provided to us by the respondents in the interviews. We did not focus on any data specific to individuals or vulnerable groups. Therefore, we do not foresee that the proposal and development of PRAC³ will cause ethical harm.

However, we do note some potential ethical issues. The coverage of our interview subjects is not comprehensive in the context of geography. Although further research is still in progress, at present, our interviewees are all from the United States. Although our summary of voice actor workflows, classification methods, and interview guidelines does not focus on regionally specific practices, and we try to make the process compatible with a more diverse AI governance environment, we cannot completely avoid this bias.

In addition, because our conclusions are mainly derived from interview data, we may have overlooked some undisclosed practices. Therefore, we emphasize that PRAC³ is a conceptual framework and may limit the comprehensive understanding of voice actors, harm the interests of vulnerable groups. 

On balance, we hope that the PRAC³ framework, developed by a cross-disciplinary team, will increase awareness of the potential risks and harms faced by voice actors and benefit a wider range of stakeholders.

\section{Positionality Statement}
All authors are currently affiliated with US academic institutions, and all our interviewees also live in the United States. None of the team members identify as a member of the voice acting community. The research team includes members who have long been engaged in risk and privacy security research, Human Computer Interaction, as well as members with extensive speech and audio technology research experience, which ensures that we are well-positioned to frame this research to understand the problems faced by voice actors from multiple perspectives.

\section{Adverse Impact Statements}
As the paper only contains non-experimental and non-identifiable interview data, we do not expect that the dissemination of this paper will have a substantial adverse impact. Our main concerns about adverse impacts are related to potential misunderstandings about the limitations of the research subjects, as described in the ethic statement.

\bigskip
%\noindent Thank you for reading these instructions carefully. We look forward to receiving your electronic files!

\bibliography{aaai25}

\end{document}